\documentclass{ws-procs9x6}
\normalsize

\begin{document}

\title{The Gluon Spin Asymmetry as a Link to $\Delta G$ and 
Orbital Angular Momentum}

\author{GORDON P. RAMSEY}

\address{Physics Dept., Loyola University Chicago and \\
HEP Division, Argonne National Lab, IL, USA \\ 
E-mail: gpr@hep.anl.gov}

\maketitle

\abstracts{
The fundamental program in high energy spin physics focuses
on the spin structure of the nucleon. The gluon and orbital 
angular momentum components of the nucleon spin are virtually 
unknown. The $J_z=\frac{1}{2}$ sum rule involves the integrated 
parton densities and can be used to extract information on the orbital angular momentum and its evolution. To avoid any bias on a model of $\Delta G$, we assume that the gluon asymmetry, $A=\Delta G/G$ can be used to extract $\Delta G$ over a reasonable kinematic region. Combining the results for 
$\Delta G$ with the evolution equations, we can determine a theoretical expression for the orbital angular momentum and its evolution.}

\section{Introduction}
The $J_z=\frac{1}{2}$ sum rule involves the integrated 
densities. 
\begin{equation}
J_z\equiv \frac{1}{2}\Delta \Sigma+\Delta G+(L_z)_q+(L_z)_G,
\end{equation}
where $\Delta \Sigma$ is the total spin carried by all quarks,
$\Delta G$ the spin carried by gluons and $(L_z)_q$ and
$(L_z)_G$ the orbital angular momenta of the quarks and
gluons respectively. Recent DIS experiments have narrowed the 
quark spin contribution ($\Delta \Sigma$) to within a 
reasonable degree. However, the gluon and orbital angular momentum components of the nucleon spin are virtually unknown. 
The purpose of this work is to estimate the gluon density
through use of the asymmetry $A=\Delta G/G$ and, with the 
evolution equations, determine a theoretical expression 
for the orbital angular momentum.

There are numerous models for the polarized quark distributions,
most of which are in agreement with the polarized DIS data.
We have chosen to use the model of Gordon, et. al.\cite{ggr},
which separates the valence and sea flavors and builds in an
asymmetry of the quark and antiquark distributions. These
polarized distributions depend upon the corresponding 
unpolarized distributions. To investigate the effects of a 
range of polarized PDFs, both CTEQ5\cite{cteq} and 
MRST2001\cite{mrst} unpolarized distributions are used. 
For the distributions considered, the x-dependence of the quark 
spin content is identical above x=0.10, but varies considerably 
for smaller x. Similarly, the different models for the 
unpolarized glue lead to a few percent differences in the gluon
asymmetry and the $x$ dependent values of $L_z$ at small-$x$. 

Our present knowledge of the polarized flavor distributions 
comes mostly from polarized deep-inelastic-scattering (PDIS).
The up and down quark polarizations are fairly well established,
but the strange and charm quark distributions are less known.
Since these heavier quarks do not contribute more than a few
percent to the proton spin, the overall quark spin, $\Delta
\Sigma$ is known to within a few percent. Most of the accepted
values lie well within the range $0.20-0.35$. The integrated 
polarized gluon distribution is virtually unknown, with 
estimates from almost zero to somewhat large values, around 2.0.
This work consists of two parts. The first is to extract
information about $\Delta G$ from the polarized gluon 
asymmetry, $A=\Delta G/G$. This minimizes the bias of 
constructing an arbitrary model of $\Delta G$. The assumptions 
of the asymmetry are based upon sound theoretical grounds and
are directly verifiable by present experiments at DESY and
RHIC at BNL. From the asymmetry and knowledge of the unpolarized
gluon distributions (extracted from CTEQ and MRST to investigate
the variation) we can extract the necessary information about
$\Delta G$. Coupled with the $J_z=\frac{1}{2}$ sum rule, we
can then infer information about $L_z$ and its evolution in
$Q^2$.

\section{Asymmetry Model for $\Delta G$ and $L_z$}
Experiments are underway and planned to measure both $\Delta G$ 
and the polarized gluon asymmetry, $A=\Delta G/G$. To construct
a theoretical model of the asymmetry, we assume that it has a 
scale independent part, $A_0$ plus a small piece that 
vanishes at some large scale so that $\Delta G$ can be written 
as $\Delta G=A_0\cdot G+G_\epsilon$. The scale invariant $A_0$ 
is calculable and is independent of theoretical models of 
$\Delta G$. The second term is scale invariant and is 
interpreted as the difference between the measured polarized 
gluon distribution and that predicted by the calculated $A_0$ 
combined with measurement of the unpolarized gluon density. 
Experimental results can be correlated by measuring the 
asymmetry, $A$ in a limited kinematic range of $x$ for a fixed 
$Q^2$ (at HERMES and RHIC). The invariance of $A_0$ and 
$G_\epsilon$ can then be used to extract $\Delta G$ over an 
expanded kinematic region. Combining this with the extraction 
of $\Delta G$ in polarized $pp$ collisions at RHIC can enhance 
the correlations of these experimental results. Thus, the model
for $A$ can be readily verified in separate sets of experiments.

We can write the asymmetry as 
\begin{equation}
A(x,t)=A_0(x)+\epsilon(x,t)\equiv \Delta G/G 
\end{equation}
where $t\equiv \ln[(\alpha_s(Q_0^2)/(\alpha_s(Q^2)]$.
Then, $\Delta G$ can be written in terms of the calculated 
asymmetry $A_0(x)$ and a difference term. The calculable part
can be found by taking the $t$-derivative of $A_0(x)$:
\begin{equation}
\frac{dA_0(x)}{dt}=\frac{\Delta G}{dt}-A_0\cdot \frac{dG}{dt}=0
\end{equation}
to a first approximation. Then $\frac{d\Delta G}{dt}$ and 
$\frac{dG}{dt}$ are calculated using the evolution equations. 
The result is a simple polynomial in $x$. The quantity 
$\epsilon(x,t)\cdot G(x,t)$ is scale-invariant at some large scale so $\frac{d\epsilon}{dt}$ is also calculable. 
From the counting rules, we bound $\epsilon(x,t)$ by 
$\epsilon(x,t)\leq c(t)\cdot x(1-x)$. We require $\epsilon(x,t)$ to be decreasing at some scale, since 
$\epsilon(x,t)\cdot G(x,t)$ is scale invariant. Then, the form 
for $\epsilon(x,t=0)=x(1-x)^n$, where $n$ is the power of 
$(1-x)$ in $G(x)$ and we assume that $c(t)\equiv 1$. Then, the
calculation of $L_z$ and its evolution involves using the 
$J_z=\frac{1}{2}$ sum rule and the DGLAP evolution equations.
\begin{equation}
L_z=\frac{1}{2}-\Delta \Sigma/2-(A_0+\epsilon)\cdot G. 
\end{equation}
From its derivative with respect to $t$ and the evolution
equations, the evolution of $L_z$ is
\begin{equation}
\frac{dL_z}{dt}=-[\Delta P_{qq}\otimes \Delta q+\Delta P_{qG}\otimes (A_0\cdot G)]/2-A_0[P_{Gq}\otimes q+P_{GG}\otimes G].
\end{equation}

\section{Results and Experimental Verification}
A plot of the $L_z(x)$ is shown in figure 1. Differences in
the MRS and CTEQ based distributions can be seen at small-$x$.
The evolved $L_z$ for the CTEQ based model is also shown in the 
figure as a dotted line. The evolution to $100$ GeV$^2$ is
not that significant at leading order.

The model of the gluon asymmetry $A$ can be tested by three
separate experiments at DESY (HERA), BNL (RHIC) and CERN 
(COMPASS). First, $\Delta G$ can be measured via prompt photon 
production or jet production. These processes yield the largest 
asymmetries for the size of $\Delta G$\cite{gr}. Also 
measurement of $\Delta G/G$ in at least a small kinematic range 
of $x$ and $Q^2$ will test this model.\cite{exps} The 
corresponding predictions for the 
orbital angular momentum $L_z$ and its evolution can be 
measured via deeply-virtual Compton Scattering (DVCS) and 
vector-meson scattering. 

We have proposed a way to calculate the orbital angular 
momentum and its evolution through use of the gluon asymmetry,
$A=\Delta G/G$. Experiments are outlined here to test the 
various assumptions and results obtained in this model.

\begin{figure}
%\epsfxsize=10cm   
%width of figure - will enlarge/reduce the figures
\rotatebox{270}{\resizebox{3.0in}{!}{\includegraphics{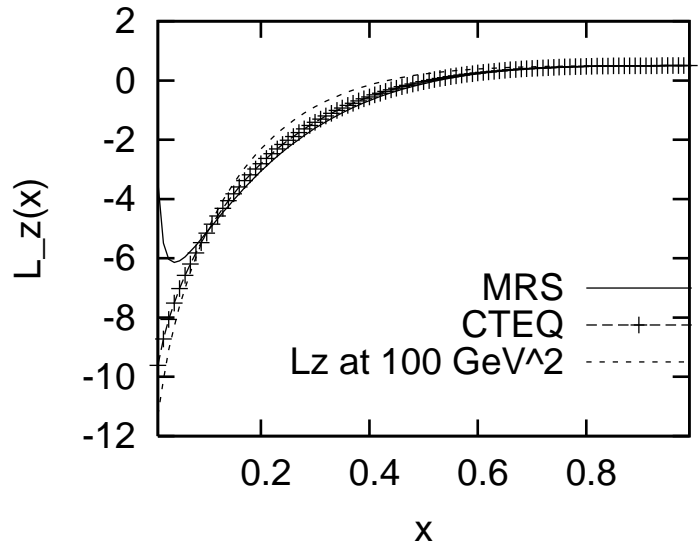}}}
%\figurebox{10cm}{8cm}{lz} %to have a box alone 
%{\epsfxsize=4in\epsfbox{lz.eps}}}
\caption{Orbital angular momentum versus x generated by MRS 
(lines) and CTEQ (crosses) distributions and evolved to 
$Q^2=100$ GeV$^2$ squared (dashes). \label{lzplot}}
\end{figure}
%Next adjust the scaling of the figure until it is correctly 
%positioned, and remove the declarations of the lines and any 
%anomalous spacing. The caption heading for a figure should be 
%placed below that figure.

\section*{Acknowledgments}
This work is supported in part by the U.S. Department of
Energy, Division of High Energy Physics, Contract 
W-31-109-ENG-38.


\begin{thebibliography}{0}
\bibitem{ggr} L. Gordon, M. Goshtasbpour and G. P. Ramsey, {\it 
Phys. Rev.} {\bf D58}, 094017 (1998).

\bibitem{cteq} G. Pang and H. Zhao, {\it Phys. Rev.} {\bf D65}, 
014012 (2003)

\bibitem{mrst} A. D. Martin, {\it et. al.,}{\it Eur. Phys. J.} {\bf C23}, 73 (2001).

\bibitem{gr} L. Gordon and G. P. Ramsey, {\it Phys. Rev.} {\bf D59}, 074018 (1999).

\bibitem{exps} See papers by A. Metz, C. Schill, J. Sowinski
and D. Hasch in this proceedings.

\end{thebibliography}
\end{document}